\documentclass[a4paper, amsfonts, amssymb, amsmath, reprint, showkeys, nofootinbib, twoside]{revtex4-1}

\def\prd{{Phys.~Rev.~D}}        
\def\prl{{Phys.~Rev.~Lett.}}    
\usepackage{amsmath,amstext}
\usepackage[T1]{fontenc}
\usepackage{amssymb}
\input{epsf}
\usepackage{graphicx}
\usepackage{ae,aecompl}

\usepackage{hyperref}
\usepackage{amsmath}
\usepackage{amssymb}
\usepackage{mathtools}
\usepackage{bm}
\usepackage{cleveref}
\usepackage{tensor}
\usepackage{braket}
\usepackage{enumitem}
\usepackage{mhchem}

\usepackage{color}

\def\be{\begin{equation}}
\def\ee{\end{equation}}
\def\beq{\begin{eqnarray}}
\def\eeq{\end{eqnarray}}

\begin{document}
\title{Should Unstable Quantum Field Theories be Lorentz Invariant?}
\author{L.~Gavassino}
\affiliation{Department of Mathematics, Vanderbilt University, Nashville, TN, USA}

\begin{abstract}
An unstable field theory is what we obtain when we linearise the equations of an interacting field theory near an unstable state. Theories of this kind are adopted to model the onset of spontaneous symmetry breakings, when the fields are sitting on the top of the Mexican hat, and they start to ``roll down'' to the bottom. At present, there exists no rigorous proof that unstable quantum field theories are Lorentz-invariant (in the sense of Wigner's theorem). Here, we show that they shouldn't be. In fact, unstable theories always have a limited regime of applicability, and they are valid only for a very short time. As consequence, there is a preferred simultaneity hyperplane, along which the unstable theory is everywhere applicable, while a generic observer (whose four-velocity is not orthogonal to such hyperplane) must use the full non-linear theory.
In summary: the current quantization schemes are ``ok'', independently from whether they lead to a Lorentz-invariant theory.
\end{abstract} 

\maketitle

\section{Introduction}

In the simplest model of spontaneous symmetry breaking \cite{felder2001,TachyonicSymmetryBr2001}, one has a real scalar field $\varphi$, whose dynamics is governed by the Lagrangian density\footnote{We adopt the metric signature $(-,+,+,+)$, and work in natural units: $c=\hbar=1$.}
\begin{equation}\label{lagro}
\mathcal{L} = -\dfrac{1}{2} \partial_\mu \varphi \, \partial^\mu \varphi -\dfrac{\lambda}{4}(\varphi^2-v^2)^2 \, , 
\end{equation}
with $\lambda,v>0$. Here, the symmetry that is being broken is the reflection $\varphi \rightarrow -\varphi$, as the system needs to choose one of two possible ground states: $\varphi=\pm v$. The Euler-Lagrange equation for the Lagrangian density \eqref{lagro} is
\begin{equation}\label{nonlinearizzo}
\partial_\mu \partial^\mu \varphi = -m^2 \varphi+\lambda \varphi^3  \, ,
\end{equation}
with $m^2=\lambda v^2$. Now, suppose that the initial value of the field is close to zero. Since the state $\varphi=0$ is a local maximum of the potential, the system is in a unstable state, and it will start ``rolling down'' towards the bottom of the potential (i.e. towards $\varphi=\pm v$). However, during the initial phase of the fall, $\varphi$ is so small that the term $\lambda \varphi^3$ is effectively negligible. Thus, we can linearise equation \eqref{nonlinearizzo}, and work with the much simpler field equation
\begin{equation}\label{linearizzo}
\partial_\mu \partial^\mu \varphi = -m^2 \varphi  \, ,
\end{equation}
which is known as the ``tachyon field equation'' \cite{Feinberg1967}. 


Here is the issue. If we want to make reliable models of spontaneous symmetry breakings, we \textit{need} a quantum theory. The reason is that, in classical field theory, the state $\varphi=0$ is an equilibrium state. This means that the field will not fall down, unless it is externally perturbed. On the other hand, in a quantum world, there are inevitably fluctuations (related to the uncertainty principle), and these fluctuations push the field down without the need of any external influence. Hence, a purely classical description is structurally incapable of grasping the initial dynamics of the symmetry breaking \cite{felder2001}. So, the question is: Can we quantise equation \eqref{linearizzo}?

The matter is still debated. If we look for plane-wave solution of \eqref{linearizzo}, of the form $\varphi \propto e^{i(kx-\omega t)}$, we get the dispersion relation $\omega=\sqrt{k^2-m^2}$ (like Klein-Gordon, but with the replacement $m\rightarrow im$). In the past, this led many authors to conclude that the elementary excitations of \eqref{linearizzo} should be interpreted as superluminal particles (``tachyons'' \cite{Feinberg1967}), whose four-momentum is spacelike. Hence, in their first attempts to quantize \eqref{linearizzo}, their main goal was to incorporate this alleged ``superluminal character'' into the quantum theory \cite{Feinberg1967,Sundarshan1968,Tanaka1960,
Kamoi1971,Dhar1968,Murphy1972}. However, it was soon realised that \eqref{linearizzo} is actually a perfectly causal equation, whose excitations travel inside the lightcone \cite{Susskind1969,FoxKuper1970,
GavassinoSuperluminal2021}, and that quantum field theory structurally forbids any kind of superluminal transmission of signals \cite{Eberhard1988,Keister1996,Peskin_book,Coleman2018,
GavassinoAsymmetry2022}. Hence, the ``superluminal interpretation'' has largely been abandoned, in favour of more conventional approaches \cite{Schroer1970,Lima2013,Efimov2012}.

One issue that immediately appears when we try to quantise unstable ``tachyon-type'' equations like \eqref{linearizzo} using standard techniques is that the resulting theory does not have a vacuum state \cite{Schroer1970}, thus violating one of the Wightman axioms \cite{wightman_book}.  \citet{Schroer1970} tried to fix this by considering a Hilbert space with indefinite metric. However, if one thinks about it, the non-existence of a vacuum state is precisely what we want. In fact, if the Klein-Gordon theory is the ``field-theory analogue'' of the harmonic oscillator (in the sense that each Fourier mode is an oscillator), the tachyon field is the analogue of the inverted oscillator \cite{Subramanyan2021}, whose Hamiltonian is
\begin{equation}
H(x,p)= \dfrac{p^2}{2m} -\dfrac{1}{2} \kappa x^2 \, .
\end{equation}
We do not expect such a system to admit a ground state, because it is unstable by construction. Of course, the ``exact'' theory arising from the Lagrangian  \eqref{lagro} admits a vacuum (actually, two vacua, with $\braket{\varphi}=\pm v$). But such vacuum states fall outside the regime of validity of the tachyon approximation \eqref{linearizzo}, which holds only for $|\varphi|\ll v$. Hence, the non-existence of a ``tachyonic'' vacuum state is to be expected \cite{Shiff1940}, and even  desired \cite{Lima2013}.

There is still, however, another issue that needs to be solved: the question about Lorentz invariance. Of course, the Lagrangian density \eqref{lagro} should give rise to a Lorentz-invariant field theory. And, indeed, the field equation \eqref{nonlinearizzo} is Lorentz-invariant: it looks the same in all reference frames. Also equation \eqref{linearizzo} is Lorentz-invariant, as it is obtained from \eqref{nonlinearizzo} under the assumption $|\varphi|\ll v$, which is a Lorentz-invariant condition (since $\varphi$ is a scalar). Hence, it is reasonable to expect that also \eqref{linearizzo} will give rise to a Lorentz-invariant theory\footnote{Indeed, in a classical world, equation \eqref{linearizzo} describes a theory which is manifestly Poincar\'{e}-invariant. Its effective Lagrangian density is a Lorentz scalar, $2\mathcal{L}=-\partial_\mu \varphi \partial^\mu \varphi -m^2 \varphi^2$, which does not depend explicitly on $x^\mu$. Thus, no preferred direction in spacetime is singled out. Another perspective on this is that, since in a classical world the condition $\varphi(p)=0$ $\forall p \in \mathbb{R}^{1+3}$ defines a Poincar\'{e}-invariant state (the so called ``false vacuum''), we are linearising a Lorentz-invariant theory around a Lorentz-invariant background state, and this must produce a Lorentz-invariant linear theory.}. But one needs to be careful, because in quantum field theory, the expression ``Lorentz invariance'' has a rigorous mathematical meaning. It is not just a statement about how the field equations look like in different reference frames. By Wigner's theorem, a quantum theory is Lorentz-invariant if and only if its Hilbert space is the carrier space of a unitary representation $U(\Lambda)$ of the Lorentz group \cite{weinbergQFT_1995,BjorkenDrell_book,Exner:1983xu}. Unfortunately, there is to date no rigorous proof that this is indeed the case, for unstable fields (like the tachyon field). For example, the quantization scheme of \citet{Lima2013} is carried out within a generic static spacetime, with no other specific symmetries besides time-translation invariance. Hence, the Minkowski limit of Lima's theory is not guaranteed to be consistent with Lorentz invariance. Indeed, the issue of whether tachyon field theories are Lorentz-invariant has been a main reason of concern since the beginning \cite{Tanaka1960,Sundarshan1968,Kamoi1971}, and the matter is still debated \cite{Radzikowski2008,Perepelitsa2016}.

Our goal, here, is to show that, if we start from the assumption that the tachyon field equation \eqref{linearizzo} is just the linear limit of equation \eqref{nonlinearizzo}, then the associated ``quantum tachyonic theory'' is not expected to be Lorentz-invariant (in Wigner's sense). Actually, even if we were able to find some operators $U(\Lambda)$ that resemble the structure of the Lorentz group, such operators would be unphysical, and their action on physical states should not be trusted.

\section{Tachyon fields are not Lorentz-invariant}

In subsection \ref{intu}, we present an intuitive geometric argument as to why tachyon field theories should not be expected to be Lorentz-invariant. In subsection \ref{rigo}, we set up an algebraic (operator-based) argument.

\subsection{Geometric argument}\label{intu}

\begin{figure*}
\begin{center}
\includegraphics[width=0.9\textwidth]{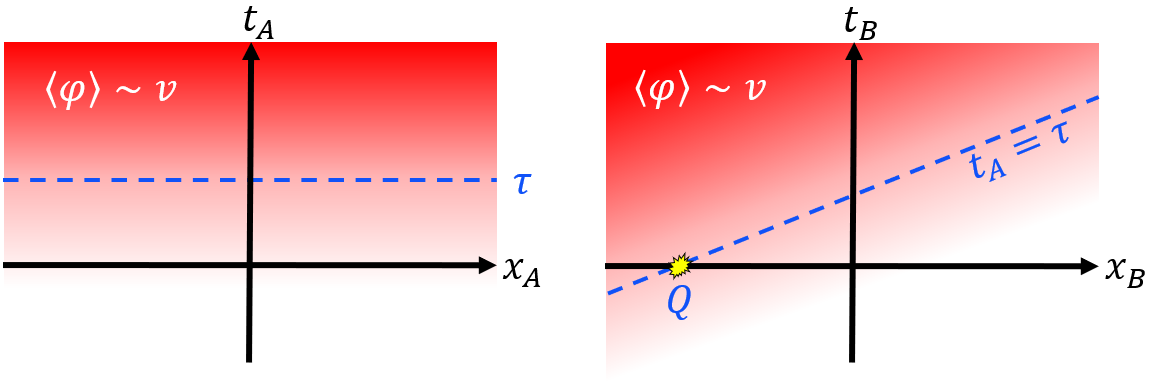}
	\caption{Minkowski diagram of $\braket{\varphi}$ in Alice's coordinates (left panel), and in Bob's coordinates (right panel). The two diagrams are mapped into each other by a Lorentz boost. The shades of red are a color-map of the intensity of $\braket{\varphi}$ (red large, white small). The blue dashed line marks the threshold above which the tachyon field approximation is no longer valid, as $\braket{\varphi} \sim v$. Above such threshold, one needs to rely on the full theory (restoring the non-linear term $\lambda \varphi^3 $). In Alice's frame, this happens at a certain time $\tau$. In Bob's frame, instead, it happens in different places at different times. Crucially, there are some locations where the tachyon approximation is not valid even at $t_B=0$. These are all the places on the left of $Q=-\tau/ \gamma w$ (yellow star).}
	\label{fig:fig}
	\end{center}
\end{figure*}

Let Alice be an inertial observer, and call ``$t_A$'' her time coordinate. Suppose that in her reference frame, at $t_A=0$, the quantity $\braket{\varphi^2}$ is close to zero, so that the tachyon approximation is valid. Then, taking the average of equation \eqref{linearizzo}, we find that the field average should grow exponentially, over a characteristic timescale $m^{-1} \,$:
\begin{equation}\label{gvbuf}
\braket{\varphi(t_A)} \sim e^{mt_A} \, .
\end{equation}
At some time $\tau$, the average of $\varphi$ becomes so large, that the tachyon approximation is no longer valid. This happens when $\braket{\varphi} \sim v $. Hence, for $t_A \gtrsim \tau$, we need to rely on the  ``exact'' theory, and we must restore the non-linear term $\lambda \varphi^3 $ in the field equation.

Now, let us see what happens in the reference of Bob, who moves with speed $w>0$ with respect to Alice. The change of coordinates that relates Alice's and Bob's frames is (if we orient the axes properly)
\begin{equation}\label{boost}
\left\{ 
\begin{array}{ll} 
t_A = \gamma (t_B -w x_B) \, ,\\
x_A = \gamma (x_B -w t_B) \, ,\\
\end{array}\right.
\end{equation} 
with $\gamma=(1-w^2)^{-1/2}$. Hence, equation \eqref{gvbuf} becomes, in Bob's coordinates (recall that $\varphi$ is a scalar),
\begin{equation}
\braket{\varphi(t_B,x_B)} \sim e^{m\gamma t_B} e^{-m\gamma wx_B} \, .
\end{equation}
As we can see, in Bob's frame the field average has an exponential dependence both on space and on time (see figure \ref{fig:fig}). This is a consequence of relativity of simultaneity \cite{special_in_gen,GavassinoSuperluminal2021,
GavassinoGiacosa2022}. But now we immediately see the problem. As $x_B \rightarrow -\infty$, the value of $\braket{\varphi}$ can become arbitrarily large. In particular, there is a point $Q$, given by
\begin{equation}
Q= -\dfrac{\tau}{\gamma w} \, ,
\end{equation}
such that, on its left, the average $\braket{\varphi}$ is comparable to $v$ already at $t_B=0$. But this implies that Bob is not allowed to use the tachyon approximation at $t_B=0$. Actually, he is not allowed to use it at any time, because for any $t_B$ there is some location $Q(t_B)$ such that the tachyon approximation breaks down on the left of $Q(t_B)$.

In practice, Bob may just focus on observables with support on the region $x_B>Q(t_B)$, where the tachyon approximation is still valid. However, the focus of Wigner's theorem is on \textit{quantum states}, which describe the totality of the system at a given time. In particular, if the tachyon theory is Lorentz-invariant, then for any quantum state (of the tachyon theory) that describes the system in Alice's frame, there must be a corresponding quantum state (belonging to the tachyon theory itself) that describes the same system in Bob's frame. Clearly, this does not happen, as the state in Bob's frame can only be described within the full non-linear theory, at any time.

Note that this result is highly non-trivial, as it arises from the intrinsic differences between the classical and the quantum descriptions of a same system. In a classical description, everything we need is a field, $\varphi:\mathbb{R}^{1+3}\rightarrow \mathbb{R}$,  which obeys a Lorentz-invariant field equation. From this perspective, equation \eqref{linearizzo} defines a theory that is manifestly Lorentz-invariant. However, in a quantum world, we also need a unitary group of transformations $U(\Lambda)$, which does not exist here.

\subsection{Algebraic argument}\label{rigo}

Let $\mathcal{H}$ be the Hilbert space of the full non-linear theory arising from the Lagrangian \eqref{lagro}. Let also
\begin{equation}
\begin{split}
 U(\Lambda) :{}& \mathcal{H} \rightarrow \mathcal{H} \, , \\
 H :{}& \mathcal{H} \rightarrow \mathcal{H} \, , \\
 P^j :{}& \mathcal{H} \rightarrow \mathcal{H} \, ,\\
\end{split}
\end{equation}
be respectively the Lorentz group, the Hamiltonian, and the linear momentum of the non-linear theory\footnote{Note that $U(\Lambda)$ here describes a Lorentz transformation associated to the full non-linear theory, and not to the tachyon theory.}. For example, the operator $H$ is just \cite{weinbergQFT_1995}
\begin{equation}
H = \dfrac{1}{2} \int \bigg[ (\partial_t \varphi)^2 + |\nabla \varphi|^2 + \dfrac{\lambda}{2}(\varphi^2-v^2)^2 \bigg] \, d^3 x \, .
\end{equation}
Let us now define a projector $\mathcal{Q}:\mathcal{H}\rightarrow \mathcal{H}$, which returns ``$1$'' if the dynamics of the state can be well approximated (at $t=0$) using the tachyon field theory, and ``$0$'' otherwise. Clearly, if a state $\ket{\alpha}$ can be described (at $t=0$) within the tachyon theory, the same must be true for $e^{-iP^ja_j}\ket{\alpha}$, which is just a copy of $\ket{\alpha}$ translated in space. Hence, $\mathcal{Q}$ is invariant under space translations:
\begin{equation}\label{MaQ}
[\mathcal{Q} ,P^j]=0 \, .
\end{equation}
On the other hand, if we evolve the state $\ket{\alpha}$ for a long time, at some point the tachyon approximation ceases to be valid, because $\braket{\varphi}$ becomes too large. Hence, 
\begin{equation}
[\mathcal{Q},H] \neq 0 \, .
\end{equation}
Now, let $\Lambda$ be a boost of velocity $w \neq 0$ in the $x^1$ direction, and let us introduce the short-hand notation $\ket{\Lambda\psi}:= U(\Lambda)\ket{\psi}$.
Our goal is to show that we can always find at least one state $\ket{\beta}$ such that
\begin{equation}\label{beTTone}
\begin{split}
& \mathcal{Q}\ket{\beta}=\ket{\beta} \, , \\
& \mathcal{Q} \ket{\Lambda\beta} \approx 0 \, . \\
\end{split}
\end{equation}  
That is, we can always find a state $\ket{\beta}$ that can be modelled within the ``tachyon approximation'', but such that, when we boost it, the resulting state $\ket{\Lambda\beta}$ cannot be described within the tachyon theory itself. The existence of such a state would imply that the tachyon field theory is not Lorentz-invariant (in Wigner's sense), because a boost is not a one-to-one transformation from the tachyonic Hilbert space to itself (not even approximately!).

Pick a quantum state $\ket{\alpha}$ such that $\mathcal{Q}\ket{\alpha}=\ket{\alpha}$, and consider the one-parameter family of states $\ket{\alpha(L)}=e^{iP^1L}\ket{\alpha}$. These are all eigenstates of $\mathcal{Q}$ with eigenvalue 1, since
\begin{equation}
\mathcal{Q}\, e^{iP^1L}\ket{\alpha}=e^{iP^1L}\mathcal{Q}\ket{\alpha}=e^{iP^1L}\ket{\alpha} \, .
\end{equation}
On the other hand, we can use the composition rules of the Poincar\'e group \cite{weinbergQFT_1995} to show that \cite{GavassinoGiacosa2022}
\begin{equation}\label{Poincoro}
U(\Lambda) \, e^{iP^1L}\ket{\alpha} = e^{iP^1\gamma L}e^{-iH\gamma w L}U(\Lambda)\ket{\alpha} \, .
\end{equation}
Hence, recalling equation \eqref{MaQ}, we have
\begin{equation}\label{qqqqq}
\mathcal{Q}\ket{\Lambda\alpha(L)}=e^{iP^1\gamma L}\mathcal{Q}e^{-iH\gamma wL}\ket{\Lambda \alpha} \, .
\end{equation}
Now, let us focus on the time-evolution of $\ket{\Lambda\alpha}$. As we said before, we expect that all states, sooner or later (as the field ``rolls down''), will exit the tachyonic approximation. Hence, there should be one instant of time $\tau$ such that $\mathcal{Q}e^{-iH\tau}\ket{\Lambda\alpha}\approx 0$. Therefore, if we choose $L=\tau/w\gamma$ (for this we need $w \neq 0$), equation \eqref{qqqqq} finally becomes
\begin{equation}
\mathcal{Q}\ket{\Lambda\alpha(\tau/w\gamma)}=e^{iP^1\gamma L}\mathcal{Q}e^{-iH\tau}\ket{\Lambda \alpha} \approx 0 \, .
\end{equation}
Thus, if we set $\ket{\beta}=\ket{\alpha(\tau/w\gamma)}$, we recover both equations in \eqref{beTTone}. This completes our proof.


Let us make some quick final remarks:
\begin{itemize}
\item Also the algebraic argument above makes heavy use of relativity of simultaneity. In fact, the evolution operator $e^{-iH\gamma wL}$ on the right-hand side of \eqref{Poincoro} comes from the fact that, since $e^{iP^1L}\ket{\alpha}$ and $\ket{\alpha}$ have a ``distance'' $L$, when we boost them, their internal clocks are desynchronized of an amount $\gamma w L$. 
\item To restore Lorentz-invariance, one may try to artificially remove from the tachyonic Hilbert space all those states $\ket{\beta}$ that satisfy \eqref{beTTone}. But this would imply that for any tachyonic state $\ket{\alpha}$, there exists a value of $L$ such that $e^{iP^1L}\ket{\alpha}$ is not tachyonic. Hence, to restore Lorentz-invariance, we would end up breaking space-translation invariance. 
\item We did not make any assumption about the magnitude of $w$, besides $w \neq 0$. Hence, Lorentz-invariance is broken at any non-vanishing speed.
\end{itemize}

\section{Conclusions}

Here is the good news. The quantization schemes of unstable theories that are available in the literature  should be considered acceptable even if a rigorous analysis were to show that the Hilbert space is not equipped with a unitary representation of the Lorentz group. The problem, in a nutshell, is that the regime of validity of an unstable theory necessarily lasts for a finite amount of time. Sooner or later, we must replace the unstable theory with the full non-linear theory. Then, because of relativity of simultaneity, if the ``tachyon-like'' approximation is valid across all space (at a given time) in Alice's frame, it \textit{cannot} be valid across all space also in Bob's frame [see figure \ref{fig:fig}].

All of this does not mean that we should also observe a breakdown of Lorentz-invariance in actual computations. Equation \eqref{linearizzo} is still a Lorentz-invariant field equation. Hence, there can be no trace of this ``broken Lorentz symmetry'' in the dynamics of $\braket{\varphi}$. Additionally, since equation \eqref{linearizzo} is linear, the field-field commutator $[\varphi(x),\varphi(y)]$ obeys the same tachyon equation as the field itself, and such equation is Lorentz-invariant. Hence, also the commutator should not give any signal of a broken Lorentz symmetry. Indeed, no ``local'' computation should be affected by these issues. The problem appears only at a global level, when we look at the Hilbert space as whole, and we consider phenomena that take place at spacelike infinity ($x^1 \rightarrow \infty$), where the effects of relativity of simultaneity become divergent ($t'=\gamma t-\gamma w x^1 \rightarrow \infty$).

From a mathematical perspective, our analysis is similar to that carried out in a previous article \cite{GavassinoGiacosa2022}, where it was shown that whether an unstable particle has decayed or not may depend on the reference frame, if relativity of simultaneity is taken into account. Although the physical setting is different, there is a common moral to both works: when an unstable system of any kind is Lorentz-boosted, relativity of simultaneity can give rise to many counterintuitive effects \cite{GavassinoSuperluminal2021,
GavassinoLyapunov_2020,GavassinoUEIT2021}, which have been overlooked till now.

\section*{Acknowledgements}

This work was supported by a Vanderbilt's Seeding Success Grant. I thank F. Giacosa for reading the manuscript and providing useful comments.

\bibliography{Biblio}

\begin{thebibliography}{31}%
\makeatletter
\providecommand \@ifxundefined [1]{%
 \@ifx{#1\undefined}
}%
\providecommand \@ifnum [1]{%
 \ifnum #1\expandafter \@firstoftwo
 \else \expandafter \@secondoftwo
 \fi
}%
\providecommand \@ifx [1]{%
 \ifx #1\expandafter \@firstoftwo
 \else \expandafter \@secondoftwo
 \fi
}%
\providecommand \natexlab [1]{#1}%
\providecommand \enquote  [1]{``#1''}%
\providecommand \bibnamefont  [1]{#1}%
\providecommand \bibfnamefont [1]{#1}%
\providecommand \citenamefont [1]{#1}%
\providecommand \href@noop [0]{\@secondoftwo}%
\providecommand \href [0]{\begingroup \@sanitize@url \@href}%
\providecommand \@href[1]{\@@startlink{#1}\@@href}%
\providecommand \@@href[1]{\endgroup#1\@@endlink}%
\providecommand \@sanitize@url [0]{\catcode `\\12\catcode `\$12\catcode
  `\&12\catcode `\#12\catcode `\^12\catcode `\_12\catcode `\%12\relax}%
\providecommand \@@startlink[1]{}%
\providecommand \@@endlink[0]{}%
\providecommand \url  [0]{\begingroup\@sanitize@url \@url }%
\providecommand \@url [1]{\endgroup\@href {#1}{\urlprefix }}%
\providecommand \urlprefix  [0]{URL }%
\providecommand \Eprint [0]{\href }%
\providecommand \doibase [0]{http://dx.doi.org/}%
\providecommand \selectlanguage [0]{\@gobble}%
\providecommand \bibinfo  [0]{\@secondoftwo}%
\providecommand \bibfield  [0]{\@secondoftwo}%
\providecommand \translation [1]{[#1]}%
\providecommand \BibitemOpen [0]{}%
\providecommand \bibitemStop [0]{}%
\providecommand \bibitemNoStop [0]{.\EOS\space}%
\providecommand \EOS [0]{\spacefactor3000\relax}%
\providecommand \BibitemShut  [1]{\csname bibitem#1\endcsname}%
\let\auto@bib@innerbib\@empty
\bibitem [{\citenamefont {{Felder}}\ \emph {et~al.}(2001)\citenamefont
  {{Felder}}, \citenamefont {{Garc{\'\i}a-Bellido}}, \citenamefont {{Greene}},
  \citenamefont {{Kofman}}, \citenamefont {{Linde}},\ and\ \citenamefont
  {{Tkachev}}}]{felder2001}%
  \BibitemOpen
  \bibfield  {author} {\bibinfo {author} {\bibfnamefont {G.}~\bibnamefont
  {{Felder}}}, \bibinfo {author} {\bibfnamefont {J.}~\bibnamefont
  {{Garc{\'\i}a-Bellido}}}, \bibinfo {author} {\bibfnamefont {P.~B.}\
  \bibnamefont {{Greene}}}, \bibinfo {author} {\bibfnamefont {L.}~\bibnamefont
  {{Kofman}}}, \bibinfo {author} {\bibfnamefont {A.}~\bibnamefont {{Linde}}}, \
  and\ \bibinfo {author} {\bibfnamefont {I.}~\bibnamefont {{Tkachev}}},\ }\href
  {\doibase 10.1103/PhysRevLett.87.011601} {\bibfield  {journal} {\bibinfo
  {journal} {\prl}\ }\textbf {\bibinfo {volume} {87}},\ \bibinfo {eid} {011601}
  (\bibinfo {year} {2001})},\ \Eprint {http://arxiv.org/abs/hep-ph/0012142}
  {arXiv:hep-ph/0012142 [hep-ph]} \BibitemShut {NoStop}%
\bibitem [{\citenamefont {Felder}\ \emph {et~al.}(2001)\citenamefont {Felder},
  \citenamefont {Kofman},\ and\ \citenamefont
  {Linde}}]{TachyonicSymmetryBr2001}%
  \BibitemOpen
  \bibfield  {author} {\bibinfo {author} {\bibfnamefont {G.}~\bibnamefont
  {Felder}}, \bibinfo {author} {\bibfnamefont {L.}~\bibnamefont {Kofman}}, \
  and\ \bibinfo {author} {\bibfnamefont {A.}~\bibnamefont {Linde}},\ }\href
  {\doibase 10.1103/PhysRevD.64.123517} {\bibfield  {journal} {\bibinfo
  {journal} {Phys. Rev. D}\ }\textbf {\bibinfo {volume} {64}},\ \bibinfo
  {pages} {123517} (\bibinfo {year} {2001})}\BibitemShut {NoStop}%
\bibitem [{\citenamefont {Feinberg}(1967)}]{Feinberg1967}%
  \BibitemOpen
  \bibfield  {author} {\bibinfo {author} {\bibfnamefont {G.}~\bibnamefont
  {Feinberg}},\ }\href {\doibase 10.1103/PhysRev.159.1089} {\bibfield
  {journal} {\bibinfo  {journal} {Phys. Rev.}\ }\textbf {\bibinfo {volume}
  {159}},\ \bibinfo {pages} {1089} (\bibinfo {year} {1967})}\BibitemShut
  {NoStop}%
\bibitem [{\citenamefont {Arons}\ and\ \citenamefont
  {Sudarshan}(1968)}]{Sundarshan1968}%
  \BibitemOpen
  \bibfield  {author} {\bibinfo {author} {\bibfnamefont {M.~E.}\ \bibnamefont
  {Arons}}\ and\ \bibinfo {author} {\bibfnamefont {E.~C.~G.}\ \bibnamefont
  {Sudarshan}},\ }\href {\doibase 10.1103/PhysRev.173.1622} {\bibfield
  {journal} {\bibinfo  {journal} {Phys. Rev.}\ }\textbf {\bibinfo {volume}
  {173}},\ \bibinfo {pages} {1622} (\bibinfo {year} {1968})}\BibitemShut
  {NoStop}%
\bibitem [{\citenamefont {Tanaka}(1960)}]{Tanaka1960}%
  \BibitemOpen
  \bibfield  {author} {\bibinfo {author} {\bibfnamefont {S.~J.}\ \bibnamefont
  {Tanaka}},\ }\href@noop {} {\bibfield  {journal} {\bibinfo  {journal}
  {Progress of Theoretical Physics}\ }\textbf {\bibinfo {volume} {24}},\
  \bibinfo {pages} {171} (\bibinfo {year} {1960})}\BibitemShut {NoStop}%
\bibitem [{\citenamefont {{Kamoi}}\ and\ \citenamefont
  {{Kamefuchi}}(1971)}]{Kamoi1971}%
  \BibitemOpen
  \bibfield  {author} {\bibinfo {author} {\bibfnamefont {K.}~\bibnamefont
  {{Kamoi}}}\ and\ \bibinfo {author} {\bibfnamefont {S.}~\bibnamefont
  {{Kamefuchi}}},\ }\href {\doibase 10.1143/PTP.45.1646} {\bibfield  {journal}
  {\bibinfo  {journal} {Progress of Theoretical Physics}\ }\textbf {\bibinfo
  {volume} {45}},\ \bibinfo {pages} {1646} (\bibinfo {year}
  {1971})}\BibitemShut {NoStop}%
\bibitem [{\citenamefont {Dhar}\ and\ \citenamefont
  {Sudarshan}(1968)}]{Dhar1968}%
  \BibitemOpen
  \bibfield  {author} {\bibinfo {author} {\bibfnamefont {J.}~\bibnamefont
  {Dhar}}\ and\ \bibinfo {author} {\bibfnamefont {E.~C.~G.}\ \bibnamefont
  {Sudarshan}},\ }\href {\doibase 10.1103/PhysRev.174.1808} {\bibfield
  {journal} {\bibinfo  {journal} {Phys. Rev.}\ }\textbf {\bibinfo {volume}
  {174}},\ \bibinfo {pages} {1808} (\bibinfo {year} {1968})}\BibitemShut
  {NoStop}%
\bibitem [{\citenamefont {Murphy}(1972)}]{Murphy1972}%
  \BibitemOpen
  \bibfield  {author} {\bibinfo {author} {\bibfnamefont {J.~E.}\ \bibnamefont
  {Murphy}},\ }\href {\doibase 10.1103/PhysRevD.6.426} {\bibfield  {journal}
  {\bibinfo  {journal} {Phys. Rev. D}\ }\textbf {\bibinfo {volume} {6}},\
  \bibinfo {pages} {426} (\bibinfo {year} {1972})}\BibitemShut {NoStop}%
\bibitem [{\citenamefont {Aharonov}\ \emph {et~al.}(1969)\citenamefont
  {Aharonov}, \citenamefont {Komar},\ and\ \citenamefont
  {Susskind}}]{Susskind1969}%
  \BibitemOpen
  \bibfield  {author} {\bibinfo {author} {\bibfnamefont {Y.}~\bibnamefont
  {Aharonov}}, \bibinfo {author} {\bibfnamefont {A.}~\bibnamefont {Komar}}, \
  and\ \bibinfo {author} {\bibfnamefont {L.}~\bibnamefont {Susskind}},\ }\href
  {\doibase 10.1103/PhysRev.182.1400} {\bibfield  {journal} {\bibinfo
  {journal} {Phys. Rev.}\ }\textbf {\bibinfo {volume} {182}},\ \bibinfo {pages}
  {1400} (\bibinfo {year} {1969})}\BibitemShut {NoStop}%
\bibitem [{\citenamefont {{Fox}}\ \emph {et~al.}(1970)\citenamefont {{Fox}},
  \citenamefont {{Kuper}},\ and\ \citenamefont {{Lipson}}}]{FoxKuper1970}%
  \BibitemOpen
  \bibfield  {author} {\bibinfo {author} {\bibfnamefont {R.}~\bibnamefont
  {{Fox}}}, \bibinfo {author} {\bibfnamefont {C.~G.}\ \bibnamefont {{Kuper}}},
  \ and\ \bibinfo {author} {\bibfnamefont {S.~G.}\ \bibnamefont {{Lipson}}},\
  }\href {\doibase 10.1098/rspa.1970.0093} {\bibfield  {journal} {\bibinfo
  {journal} {Proceedings of the Royal Society of London Series A}\ }\textbf
  {\bibinfo {volume} {316}},\ \bibinfo {pages} {515} (\bibinfo {year}
  {1970})}\BibitemShut {NoStop}%
\bibitem [{\citenamefont
  {{Gavassino}}(2022{\natexlab{a}})}]{GavassinoSuperluminal2021}%
  \BibitemOpen
  \bibfield  {author} {\bibinfo {author} {\bibfnamefont {L.}~\bibnamefont
  {{Gavassino}}},\ }\href {\doibase 10.1103/PhysRevX.12.041001} {\bibfield
  {journal} {\bibinfo  {journal} {Physical Review X}\ }\textbf {\bibinfo
  {volume} {12}},\ \bibinfo {eid} {041001} (\bibinfo {year}
  {2022}{\natexlab{a}})},\ \Eprint {http://arxiv.org/abs/2111.05254}
  {arXiv:2111.05254 [gr-qc]} \BibitemShut {NoStop}%
\bibitem [{\citenamefont {Eberhard}\ and\ \citenamefont
  {Ross}(1989)}]{Eberhard1988}%
  \BibitemOpen
  \bibfield  {author} {\bibinfo {author} {\bibfnamefont {P.~H.}\ \bibnamefont
  {Eberhard}}\ and\ \bibinfo {author} {\bibfnamefont {R.~R.}\ \bibnamefont
  {Ross}},\ }\href {\doibase 10.1007/BF00696109} {\bibfield  {journal}
  {\bibinfo  {journal} {Found. Phys.}\ }\textbf {\bibinfo {volume} {2}},\
  \bibinfo {pages} {127} (\bibinfo {year} {1989})}\BibitemShut {NoStop}%
\bibitem [{\citenamefont {Keister}\ and\ \citenamefont
  {Polyzou}(1996)}]{Keister1996}%
  \BibitemOpen
  \bibfield  {author} {\bibinfo {author} {\bibfnamefont {B.~D.}\ \bibnamefont
  {Keister}}\ and\ \bibinfo {author} {\bibfnamefont {W.~N.}\ \bibnamefont
  {Polyzou}},\ }\href {\doibase 10.1103/PhysRevC.54.2023} {\bibfield  {journal}
  {\bibinfo  {journal} {Phys. Rev. C}\ }\textbf {\bibinfo {volume} {54}},\
  \bibinfo {pages} {2023} (\bibinfo {year} {1996})}\BibitemShut {NoStop}%
\bibitem [{\citenamefont {Peskin}\ and\ \citenamefont
  {Schroeder}(1995)}]{Peskin_book}%
  \BibitemOpen
  \bibfield  {author} {\bibinfo {author} {\bibfnamefont {M.~E.}\ \bibnamefont
  {Peskin}}\ and\ \bibinfo {author} {\bibfnamefont {D.~V.}\ \bibnamefont
  {Schroeder}},\ }\href {http://www.slac.stanford.edu/~mpeskin/QFT.html} {\emph
  {\bibinfo {title} {{An introduction to quantum field theory}}}}\ (\bibinfo
  {publisher} {Addison-Wesley},\ \bibinfo {address} {Reading, USA},\ \bibinfo
  {year} {1995})\BibitemShut {NoStop}%
\bibitem [{\citenamefont {Coleman}(2018)}]{Coleman2018}%
  \BibitemOpen
  \bibfield  {author} {\bibinfo {author} {\bibfnamefont {S.}~\bibnamefont
  {Coleman}},\ }\href {\doibase 10.1142/9371} {\emph {\bibinfo {title}
  {{Lectures of Sidney Coleman on Quantum Field Theory}}}},\ edited by\
  \bibinfo {editor} {\bibfnamefont {B.~G.-g.}\ \bibnamefont {Chen}}, \bibinfo
  {editor} {\bibfnamefont {D.}~\bibnamefont {Derbes}}, \bibinfo {editor}
  {\bibfnamefont {D.}~\bibnamefont {Griffiths}}, \bibinfo {editor}
  {\bibfnamefont {B.}~\bibnamefont {Hill}}, \bibinfo {editor} {\bibfnamefont
  {R.}~\bibnamefont {Sohn}}, \ and\ \bibinfo {editor} {\bibfnamefont {Y.-S.}\
  \bibnamefont {Ting}}\ (\bibinfo  {publisher} {WSP},\ \bibinfo {address}
  {Hackensack},\ \bibinfo {year} {2018})\BibitemShut {NoStop}%
\bibitem [{\citenamefont
  {{Gavassino}}(2022{\natexlab{b}})}]{GavassinoAsymmetry2022}%
  \BibitemOpen
  \bibfield  {author} {\bibinfo {author} {\bibfnamefont {L.}~\bibnamefont
  {{Gavassino}}},\ }\href {\doibase 10.1088/1361-6382/ac9107} {\bibfield
  {journal} {\bibinfo  {journal} {Classical and Quantum Gravity}\ }\textbf
  {\bibinfo {volume} {39}},\ \bibinfo {eid} {215010} (\bibinfo {year}
  {2022}{\natexlab{b}})}\BibitemShut {NoStop}%
\bibitem [{\citenamefont {Schroer}\ and\ \citenamefont
  {Swieca}(1970)}]{Schroer1970}%
  \BibitemOpen
  \bibfield  {author} {\bibinfo {author} {\bibfnamefont {B.}~\bibnamefont
  {Schroer}}\ and\ \bibinfo {author} {\bibfnamefont {J.~A.}\ \bibnamefont
  {Swieca}},\ }\href {\doibase 10.1103/PhysRevD.2.2938} {\bibfield  {journal}
  {\bibinfo  {journal} {Phys. Rev. D}\ }\textbf {\bibinfo {volume} {2}},\
  \bibinfo {pages} {2938} (\bibinfo {year} {1970})}\BibitemShut {NoStop}%
\bibitem [{\citenamefont {{Lima}}(2013)}]{Lima2013}%
  \BibitemOpen
  \bibfield  {author} {\bibinfo {author} {\bibfnamefont {W.~C.~C.}\
  \bibnamefont {{Lima}}},\ }\href {\doibase 10.1103/PhysRevD.88.124005}
  {\bibfield  {journal} {\bibinfo  {journal} {\prd}\ }\textbf {\bibinfo
  {volume} {88}},\ \bibinfo {eid} {124005} (\bibinfo {year} {2013})},\ \Eprint
  {http://arxiv.org/abs/1309.2004} {arXiv:1309.2004 [gr-qc]} \BibitemShut
  {NoStop}%
\bibitem [{\citenamefont {{Efimov}}(2012)}]{Efimov2012}%
  \BibitemOpen
  \bibfield  {author} {\bibinfo {author} {\bibfnamefont {G.~V.}\ \bibnamefont
  {{Efimov}}},\ }\href@noop {} {\bibfield  {journal} {\bibinfo  {journal}
  {arXiv e-prints}\ ,\ \bibinfo {eid} {arXiv:1202.2757}} (\bibinfo {year}
  {2012})},\ \Eprint {http://arxiv.org/abs/1202.2757} {arXiv:1202.2757
  [hep-ph]} \BibitemShut {NoStop}%
\bibitem [{\citenamefont {Streater}\ and\ \citenamefont
  {Wightman}(1964)}]{wightman_book}%
  \BibitemOpen
  \bibfield  {author} {\bibinfo {author} {\bibfnamefont {R.~F.}\ \bibnamefont
  {Streater}}\ and\ \bibinfo {author} {\bibfnamefont {A.~S.}\ \bibnamefont
  {Wightman}},\ }\href@noop {} {\emph {\bibinfo {title} {PCT, spin and
  statistics, and all that}}}\ (\bibinfo  {publisher} {Princeton University
  Press},\ \bibinfo {year} {1964})\BibitemShut {NoStop}%
\bibitem [{\citenamefont {{Subramanyan}}\ \emph {et~al.}(2021)\citenamefont
  {{Subramanyan}}, \citenamefont {{Hegde}}, \citenamefont {{Vishveshwara}},\
  and\ \citenamefont {{Bradlyn}}}]{Subramanyan2021}%
  \BibitemOpen
  \bibfield  {author} {\bibinfo {author} {\bibfnamefont {V.}~\bibnamefont
  {{Subramanyan}}}, \bibinfo {author} {\bibfnamefont {S.~S.}\ \bibnamefont
  {{Hegde}}}, \bibinfo {author} {\bibfnamefont {S.}~\bibnamefont
  {{Vishveshwara}}}, \ and\ \bibinfo {author} {\bibfnamefont {B.}~\bibnamefont
  {{Bradlyn}}},\ }\href {\doibase 10.1016/j.aop.2021.168470} {\bibfield
  {journal} {\bibinfo  {journal} {Annals of Physics}\ }\textbf {\bibinfo
  {volume} {435}},\ \bibinfo {eid} {168470} (\bibinfo {year} {2021})},\ \Eprint
  {http://arxiv.org/abs/2012.09875} {arXiv:2012.09875 [cond-mat.mes-hall]}
  \BibitemShut {NoStop}%
\bibitem [{\citenamefont {Schiff}\ \emph {et~al.}(1940)\citenamefont {Schiff},
  \citenamefont {Snyder},\ and\ \citenamefont {Weinberg}}]{Shiff1940}%
  \BibitemOpen
  \bibfield  {author} {\bibinfo {author} {\bibfnamefont {L.~I.}\ \bibnamefont
  {Schiff}}, \bibinfo {author} {\bibfnamefont {H.}~\bibnamefont {Snyder}}, \
  and\ \bibinfo {author} {\bibfnamefont {J.}~\bibnamefont {Weinberg}},\ }\href
  {\doibase 10.1103/PhysRev.57.315} {\bibfield  {journal} {\bibinfo  {journal}
  {Phys. Rev.}\ }\textbf {\bibinfo {volume} {57}},\ \bibinfo {pages} {315}
  (\bibinfo {year} {1940})}\BibitemShut {NoStop}%
\bibitem [{\citenamefont {Weinberg}(1995)}]{weinbergQFT_1995}%
  \BibitemOpen
  \bibfield  {author} {\bibinfo {author} {\bibfnamefont {S.}~\bibnamefont
  {Weinberg}},\ }\href {\doibase 10.1017/CBO9781139644167} {\emph {\bibinfo
  {title} {The Quantum Theory of Fields}}},\ Vol.~\bibinfo {volume} {1}\
  (\bibinfo  {publisher} {Cambridge University Press},\ \bibinfo {year}
  {1995})\BibitemShut {NoStop}%
\bibitem [{\citenamefont {Bjorken}\ and\ \citenamefont
  {Drell}(1965)}]{BjorkenDrell_book}%
  \BibitemOpen
  \bibfield  {author} {\bibinfo {author} {\bibfnamefont {J.~D.}\ \bibnamefont
  {Bjorken}}\ and\ \bibinfo {author} {\bibfnamefont {S.~D.}\ \bibnamefont
  {Drell}},\ }\href@noop {} {\emph {\bibinfo {title} {Relativistic Quantum
  Fields}}}\ (\bibinfo  {publisher} {McGraw-Hill Book Company},\ \bibinfo
  {year} {1965})\BibitemShut {NoStop}%
\bibitem [{\citenamefont {Exner}(1983)}]{Exner:1983xu}%
  \BibitemOpen
  \bibfield  {author} {\bibinfo {author} {\bibfnamefont {P.}~\bibnamefont
  {Exner}},\ }\href {\doibase 10.1103/PhysRevD.28.2621} {\bibfield  {journal}
  {\bibinfo  {journal} {Phys. Rev. D}\ }\textbf {\bibinfo {volume} {28}},\
  \bibinfo {pages} {2621} (\bibinfo {year} {1983})}\BibitemShut {NoStop}%
\bibitem [{\citenamefont {{Radzikowski}}(2008)}]{Radzikowski2008}%
  \BibitemOpen
  \bibfield  {author} {\bibinfo {author} {\bibfnamefont {M.~J.}\ \bibnamefont
  {{Radzikowski}}},\ }\href@noop {} {\bibfield  {journal} {\bibinfo  {journal}
  {arXiv e-prints}\ ,\ \bibinfo {eid} {arXiv:0804.4534}} (\bibinfo {year}
  {2008})},\ \Eprint {http://arxiv.org/abs/0804.4534} {arXiv:0804.4534
  [math-ph]} \BibitemShut {NoStop}%
\bibitem [{\citenamefont {{Perepelitsa}}(2016)}]{Perepelitsa2016}%
  \BibitemOpen
  \bibfield  {author} {\bibinfo {author} {\bibfnamefont {V.~F.}\ \bibnamefont
  {{Perepelitsa}}},\ }\href@noop {} {\bibfield  {journal} {\bibinfo  {journal}
  {arXiv e-prints}\ ,\ \bibinfo {eid} {arXiv:1605.03425}} (\bibinfo {year}
  {2016})},\ \Eprint {http://arxiv.org/abs/1605.03425} {arXiv:1605.03425
  [physics.gen-ph]} \BibitemShut {NoStop}%
\bibitem [{\citenamefont {Gourgoulhon}(2013)}]{special_in_gen}%
  \BibitemOpen
  \bibfield  {author} {\bibinfo {author} {\bibfnamefont {E.}~\bibnamefont
  {Gourgoulhon}},\ }\href@noop {} {\emph {\bibinfo {title} {Special Relativity
  in General Frames: From Particles to Astrophysics}}},\ \bibinfo {edition}
  {1st}\ ed.,\ Graduate Texts in Physics\ (\bibinfo  {publisher}
  {Springer-Verlag Berlin Heidelberg},\ \bibinfo {year} {2013})\BibitemShut
  {NoStop}%
\bibitem [{\citenamefont {Gavassino}\ and\ \citenamefont
  {Giacosa}(2022)}]{GavassinoGiacosa2022}%
  \BibitemOpen
  \bibfield  {author} {\bibinfo {author} {\bibfnamefont {L.}~\bibnamefont
  {Gavassino}}\ and\ \bibinfo {author} {\bibfnamefont {F.}~\bibnamefont
  {Giacosa}},\ }\href {\doibase 10.1103/PhysRevA.106.042215} {\bibfield
  {journal} {\bibinfo  {journal} {Phys. Rev. A}\ }\textbf {\bibinfo {volume}
  {106}},\ \bibinfo {pages} {042215} (\bibinfo {year} {2022})}\BibitemShut
  {NoStop}%
\bibitem [{\citenamefont {Gavassino}\ \emph {et~al.}(2020)\citenamefont
  {Gavassino}, \citenamefont {Antonelli},\ and\ \citenamefont
  {Haskell}}]{GavassinoLyapunov_2020}%
  \BibitemOpen
  \bibfield  {author} {\bibinfo {author} {\bibfnamefont {L.}~\bibnamefont
  {Gavassino}}, \bibinfo {author} {\bibfnamefont {M.}~\bibnamefont
  {Antonelli}}, \ and\ \bibinfo {author} {\bibfnamefont {B.}~\bibnamefont
  {Haskell}},\ }\href {\doibase 10.1103/physrevd.102.043018} {\bibfield
  {journal} {\bibinfo  {journal} {Physical Review D}\ }\textbf {\bibinfo
  {volume} {102}} (\bibinfo {year} {2020}),\
  10.1103/physrevd.102.043018}\BibitemShut {NoStop}%
\bibitem [{\citenamefont {{Gavassino}}\ and\ \citenamefont
  {{Antonelli}}(2021)}]{GavassinoUEIT2021}%
  \BibitemOpen
  \bibfield  {author} {\bibinfo {author} {\bibfnamefont {L.}~\bibnamefont
  {{Gavassino}}}\ and\ \bibinfo {author} {\bibfnamefont {M.}~\bibnamefont
  {{Antonelli}}},\ }\href {\doibase 10.3389/fspas.2021.686344} {\bibfield
  {journal} {\bibinfo  {journal} {Frontiers in Astronomy and Space Sciences}\
  }\textbf {\bibinfo {volume} {8}},\ \bibinfo {eid} {92} (\bibinfo {year}
  {2021})},\ \Eprint {http://arxiv.org/abs/2105.15184} {arXiv:2105.15184
  [gr-qc]} \BibitemShut {NoStop}%
\end{thebibliography}%

\end{document}